\documentclass[twocolumn,pra,aps,showpacs,superscriptaddress,floatfix,showkeys,nofootinbib,10pt]{revtex4-2}

\usepackage{amsmath,amsfonts,amssymb,amsthm,graphicx}
\usepackage{hyperref}

\usepackage[T1]{fontenc}

\usepackage{algorithm,algpseudocode}
\usepackage{listings}

\newcommand{\ket}[1]{ | \, #1 \rangle} \newcommand{\bra}[1]{ \langle #1 \, |} 
\newcommand{\proj}[1]{\ket{#1}\bra{#1}}

\newcommand{\be}{\begin{equation}} \newcommand{\ee}{\end{equation}}
\newcommand{\ba}{\begin{aligned}} \newcommand{\ea}{\end{aligned}}

\DeclareMathOperator{\Tr}{Tr}

\begin{document}
	
	\title{Entangled Rendezvous: A Possible Application of Bell Non-Locality For Mobile Agents on Networks}
	
	\author{P.~Mironowicz} \email{piotr.mironowicz@gmail.com}
	\affiliation{Department of Algorithms and System Modeling, Faculty of Electronics, Telecommunications and Informatics, Gda\'nsk University of Technology, 80-233 Gda\'nsk, Poland}
	\affiliation{International Centre for Theory of Quantum Technologies, University of Gda\'{n}sk, Wita Stwosza 63, 80-308 Gda\'{n}sk, Poland}
	
	\date{\today}
	
	\begin{abstract}
		Rendezvous is an old problem of assuring that two or more parties, initially separated, not knowing the position of each other, and not allowed to communicate, meet without pre-agreement on the meeting point. This problem has been extensively studied in classical computer science and has vivid importance to modern applications like coordinating a fleet of drones in an enemy's territory.
		
		Quantum non-locality, like Bell inequality violation, has shown that in many cases quantum entanglement allows for improved coordination of two separated parties compared to classical sources. The non-signaling correlations in many cases even strengthened such phenomena.
		
		In this work, we analyze, how Bell non-locality can be used by asymmetric location-aware agents trying to rendezvous on a finite network with a limited number of steps. We provide the optimal solution to this problem for both agents using quantum resources, and agents with only ``classical'' computing power.
		
		Our results show that for cubic graphs and cycles it is possible to gain an advantage by allowing the agents to use assistance of entangled quantum states.
	\end{abstract}
	
	
	\maketitle
	
	Cooperation and coordination of actions are vivid factors in the achievement of common goals. Nonetheless, we all know the problems that occur in almost any case, if some communication impediments between co-working parties occur, or if the parties are supposed to coordinate their actions but cannot communicate.
	
	Paradigmatic examples of such problems are tasks from the family of rendezvous search~\cite{alpern2002rendezvous}. This covers a wide range of situations where several parties (at least two) are besteaded in some area, each not knowing where the other parties are located, and covet to meet together as soon as possible. In other words, rendezvous covers tasks of arranging a meeting point for two or more agents in some random situation when their communication (and possibly also other resources) are scarce, or even entirely devoided.
	
	The problem has been first stated in 1976 by Steve Alpern~\cite{alpern1995rendezvous}. It gained attention in the 1990s, and since then its popularity and importance have increased, together with the whole field of multi-agent distributed coordination, due to the rapid development of real-world implementations of autonomous agents, such as unmanned aerial, ground, or underwater, vehicles (including drones, that are assumed to be able to operate without human intervention), see~\cite{cao2012overview,wang2016multi,yang2021overview} for recent reviews.
	
	The problem of rendezvous belongs to classical computer science and, up to our knowledge, has not been investigated by quantum scientists, and those quantum information researchers who worked with related problems (see sec.~\ref{sec:related}) do not refer to Alpern's and others' works. One of the intentions of our paper is to bridge the scientists from both fields. For this reason, we provide a brief overview of them, so our results can be understood by classical computer scientists not knowing quantum information, and \textit{vice versa}.
	
	In the recent years it has been shown by multiple examples ~\cite{bennett1992communication,shor1994algorithms,grover1996fast,buhrman2016quantum,preskill2018quantum,zhong2020quantum} that using quantum resources can enhance solving various classical ICT tasks, referred to as \textit{quantum advantage}~\cite{preskill2012quantum}. Our contribution is in showing that an advantage can be achieved also in the rendezvous search games, where the agents, or \textit{parties}, to keep quantum information nomenclature, are allowed to use the assistance of quantum entangled states~\cite{horodecki2009quantum}.
	
	The organization of the paper is as follows. In sec.~\ref{sec:rendezvous} we overview the current state-of-the-art of rendezvous investigations, and in sec.~\ref{sec:quantum} provide a brief introduction to the formalism of quantum information used in this work. Then, in sec.~\ref{sec:related} we place our results in the context of rendezvous nomenclature, and refer to the inquires of quantum researchers that are most relevant to the topic of the gathering of many parties. In sec.~\ref{sec:methods} we describe the tools specific to our work, and in sec.~\ref{sec:results} provide the results of our calculations. We conclude and provide some lines of further research in sec.~\ref{sec:conclusions}.
	
	The work provides multiple new Bell inequalities~\cite{bell1964einstein}, but it does not aim to develop the theory of quantum nonlocality. The purpose of the paper is to show a new area of its application, that, up to our knowledge, has not been yet investigated from the point of view of quantum entanglement. Thus, the intent of this work is to show how the methods of one rich field of Bell nonlocality, can be applied to support the other rich field of research on the rendezvous problem.

	\section{Rendezvous search overview}
	\label{sec:rendezvous}
	
	Rendezvous is an example of a family of problems considered in search theory~\cite{koopman1946search,alpern2006theory,alpern2013search}.  Modern applications of rendezvous search vary from finding a common channel of communication~\cite{chang2015multichannel}, agent recharging in persistent tasks~\cite{mathew2013graph}, drones and other autonomous vehicles navigation and coordination~\cite{otto2018optimization}, or finding an astray traveler by a rescue team~\cite{leone2022search}. The results possibly concern even dating and mating strategies~\cite{mizumoto2019adaptive}.
	
	The rendezvous problem originated from the so-called \textit{search games} proposed by Isaacs in 1965~\cite{isaacs1999differential}. In those games the \textit{searcher} (or \textit{minimizer}) play a zero sum game with a \textit{hider} (or \textit{maximizer}), with the payoff given be the \textit{capture time} required for the searcher in order to find the hider.
	
	Another related problem given in 1960 by Schelling~\cite{schelling1980strategy}, was a conundrum of two parachutists who land in an area with several landmarks, and no prior agreement regarding where to meet. Schelling's game was a one-shot task, with success occurring when both parachutists decide to go to the same place. Schelling provided also a social experiment indicating that when people hope to gather in one place of a known area and without communication, they are eager to choose to head towards a location (e.g. Grand Central Station) from a small set of places named \textit{focal points}~\cite{mehta1994focal,sugden1995theory,rijt2019quest}.
	
	In the earliest formulation of the rendezvous problems, being a cooperation game (with non-zero sum), two players were assumed not to be able to distinguish between themselves, thus the problem was to find a mixed strategy to be played by both parties, designed in such a way, that the expected time required for the parties to meet was as small as possible. Such problems, where both parties are identical, are referred to in the literature as \textit{symmetric rendezvous} problems~\cite{alpern1995rendezvous,czyzowicz2019symmetry}. When parties are allowed to distinguish between themselves, the problem is called \textit{asymetric rendezvous}, and the parties (or \textit{agents}) are called \textit{anonymous}, otherwise one say that parties are labelled~\cite{alpern1995rendezvousAndGal,de2006asynchronous}.
	
	The rendezvous problem has many variants. The search space can be e.g. a graph~\cite{anderson1990rendezvous} (sometimes called a \textit{network}~\cite{kranakis2006mobile,pelc2012deterministic}), a line, a circle, or a plane. A common assumption is that all parties know the structure of the whole search space, but cannot see other parties. Some works allow the parties to know their own location in the search space, in which case the parties are called \textit{location-aware}~\cite{howard1999rendezvous,collins2010tell,collins2011synchronous,banerjee2020study,czyzowicz2020gathering}, otherwise the space is called \textit{anonymous}~\cite{alpern1995rendezvous,de2006asynchronous,kranakis2006mobile}.
	
	In symmetric problems, one often tries to find a way to break the symmetry~\cite{yu1996agent}, e.g. by using randomness (mixed strategies)~\cite{alpern1995rendezvous} or asymmetry of the search space~\cite{fraigniaud2008deterministic} or starting time~\cite{pelc2019using}. For the latter cases the task is referred to as \textit{deterministic rendezvous}~\cite{de2006asynchronous,pelc2012deterministic}.
	
	Some works consider minimization of the expected time to meet~\cite{lim1997rendezvous}, or minimization of the maximum time to meet~\cite{lim1996minimax}. One may consider a rendezvous game when the parties have bounds on the total distance each of them can travel, like in the situation of traveling in cars with a limited amount of fuel. In case when the fuel limit is large enough that the meeting is sure to occur, then the fuel consumption is attempted to be minimized~\cite{alpern1997rendezvous}. Otherwise, when there is a limit on the time or the distance allowed to be covered, some authors~\cite{alpern1999rendezvous,alpern2006rendezvous,dani2016codes,georgiou2019symmetric} have investigated the maximization of the probability that the parties will eventually meet, similarly, to the parachutists' problem. Our work is also devoted to a similar problem, with the maximization of success probability.
	
	Other resources considered include memory capacity~\cite{czyzowicz2012meet,fraigniaud2008deterministic} and computational power~\cite{yu1996agent}. Scenarios with adversaries are also investigated~\cite{fomin2021can}. In some variants more than two parties are considered~\cite{lin2003multi,lin2007multi}, that is sometimes called \textit{gathering}~\cite{cieliebak2003solving}.
	
	When the search space is discrete, the parties may be allowed to move either synchronously~\cite{kranakis2003mobile} or asynchronously~\cite{de2006asynchronous}. In some models~\cite{anderson1990rendezvous} parties can meet not only in nodes, but also when the transpose their positions, \textit{i.e.} if they respectively occupy nodes $v_1$ and $v_2$ in a step $t$, and nodes $v_2$ and $v_1$ in a step $t + 1$.
	
	Additional features that parties are sometimes allowed to use include the possibility of leaving a mark by a party at the starting node~\cite{baston2001rendezvous}, marking any node, or even leaving on nodes complex notes using whiteboards, see~\cite{pelc2012deterministic} for a review.

	\section{Quantum Correlations}
	\label{sec:quantum}
	
	The formalism of quantum mechanics established near mid of 20th Century~\cite{Dirac1947} defines quantum states as positive semi-definite operators on some Hilbert space $\mathcal{H}$ of trace $1$, and the generalized measurements (so-called positive operator-valued measures, or POVMs) as collections of positive semi-definite operators that sum to identity. A state can be shared between many parties, and each party can perform its own measurement and get a result $a$ with a probability
	\begin{equation}
		p(a) = \Tr \left( \rho M^a \right),
	\end{equation}
	where $\{M^a\}_a$ is the POVM used for the measurement, $\rho$ is a quantum state, and since $M^a$ in the multi-party case acts on a subspace of $\mathcal{H}$, an implicit identity operator on the rest of $\mathcal{H}$ is used. The POVM properties can be written as $\forall_a M^a \succeq 0$ for positive semi-definiteness, and $\sum_a M^a = \openone_{\mathcal{H}}$ for sum to identity.
	
	Now, let us consider a scenario with two parties, Alice and Bob, each acting on a subspaces $\mathcal{H}^{(A)}$ and $\mathcal{H}^{(B)}$, respectively, $\mathcal{H} = \mathcal{H}^{(A)} \otimes \mathcal{H}^{(B)}$. Alice can perform one of the measurements from a set $X$, and obtain a result from a set $A$\footnote{One may think of it as Alice's measuring device has a knob with positions labeled by elements of $X$, and display able to print a single character from the set $A$.} with POVMs $\{ \{ M^a _x\}_a \}_x$, and similarly Bob with settings $Y$ and outcomes $B$, and POVMs $\{ \{ N^b _y\}_b \}_y$. Note that $M^a_x$ and $N^b_y$ act on $\mathcal{H}^{(A)}$ and $\mathcal{H}^{(B)}$, respectively. The joint probabilities of Alice and Bob are given by
	\begin{equation}
		\label{eq:PabxyQ}
		P(a,b|x,y) = \Tr \left[ \rho \left( M^a_x \otimes N^b_y \right) \right],
	\end{equation}
	for all $a \in A$, $b \in B$, $x \in X$, and $y \in Y$. We keep the above nomenclature of \textit{settings} and \textit{outcomes} also for non-quantum probability distributions.
	
	The probability distributions possible to be obtained within quantum mechanics (without communication between Alice and Bob) can be contrasted with probabilities that are possible to reach by parties that do not use quantum resources and are not allowed to communicate. This set is often called \textit{local correlations}; if parties are given some shared random variable the probabilities are called local with a hidden variable, or \textit{LHV correlations}.
	
	To be more specific, a joint distribution $P(a,b|x,y)$ is local if there exist a pair of conditional probability distributions $P^{(A)}(a|x)$ and $P^{(B)}(b|y)$ such that
	\begin{equation}
		P(a,b|x,y) = P^{(A)}(a|x) P^{(B)}(b|y).
	\end{equation}
	The joint distribution is LHV if there exist a set $\Lambda$, a probability distribution $p(\lambda)$ over $\lambda \in \Lambda$, and a pair of sets of conditional probability distributions $\{ P_{\lambda}^{(A)}(a|x) \}$ and $\{ P_{\lambda}^{(B)}(b|y) \}$ such that
	\begin{equation}
		P(a,b|x,y) = \sum_\lambda p(\lambda) P^{(A)}(a|x) P^{(B)}(b|y).
	\end{equation}
	
	Another set of probability distributions that is often considered, are \textit{non-signalling} (NS) probabilities~\cite{popescu1994quantum}. They are required to satisfy a condition that their local marginals are well defined, \textit{i.e.}
	\begin{equation}
		P(a|x,y) \equiv \sum_b P(a,b|x,y)
	\end{equation}
	does not depend on $y$, and the same with roles of Alice and Bob swapped. This set is the largest allowed by physical theories where there is no instantaneous communication, meaning that there is no way that the choice of the setting of Alice is detected by the probability distribution of outcomes of Bob, and \textit{vice versa}. On the other hand, the probability of the outcome of one party can depend on the outcome of the other party; this phenomenon is called \textit{steering}~\cite{uola2020quantum}. Steering is present also in quantum theory but to a weaker extend~\cite{schrodinger1935discussion,wiseman2007steering,oppenheim2010uncertainty,ramanathan2018steering}.
	
	Denote by $\mathfrak{L}$, $\mathfrak{S}$, $\mathfrak{Q}$ and $\mathfrak{N}$ the sets of all possible joint probability distributions that are local, LHV, quantum, or non-signaling, respectively. Since we usually abstract from the description of how a particular joint probability is realized, as long as it belongs to the set we consider, we refer to it as a \textit{box}. One part of the box is possessed by Alice, the other by Bob, and each party has access to her or his part of the outcome.
	
	A set of coefficient $\mathbb{B} \equiv \{ B_{a,b,x,y} \} \subset \mathbb{R}$ applied to a joint distribution as
	\begin{equation}
		\label{eq:game}
		\mathbb{B}[P] \equiv \sum_{a,b,x,y} B_{a,b,x,y} P(a,b|x,y)
	\end{equation}
	is called a \textit{two-partite game}. One often considers the maximal value of~\eqref{eq:game} within a given set. In particular, let
	\begin{equation}
		\label{eq:BellIneq}
		L \equiv \max_{P \in \mathfrak{P}} \mathbb{B}[P]
	\end{equation}
	with $\mathfrak{P} = \mathfrak{L}$. It is possible to find games in which there exist $P_Q \in \mathfrak{Q}$ such that $\mathbb{B}[P_Q] > L$. Such games are referred to as \textit{Bell operators}, the formula~\eqref{eq:BellIneq} is called a \textit{Bell inequality} and $P_Q$ is said to \textit{violate} the Bell inequality~\cite{clauser1969proposed}. The upper bound~\eqref{eq:BellIneq} on the value of a game~\eqref{eq:game} within a given set of joint probabilities $\mathfrak{P}$ is called a \textit{Tsirelon's bound}~\cite{cirel1980quantum}.
	
	A necessary condition for a state $\rho$ shared by parties to allow obtaining a probability distribution that violates a Bell inequality is to be \textit{entangled}~\cite{horodecki2009quantum}, that is a quantum property not possible to be reproduced with classical resources no matter with how powerful are computational capabilities of the parties~\cite{gill2003accardi}.
	
	Optimization of values of games over $\mathfrak{L}$, $\mathfrak{S}$, and $\mathfrak{Q}$ is very difficult~\cite{navascues2007bounding,navascues2008convergent}. For $\mathfrak{Q}$ one often consider a superset of joint distributions called \textit{macroscopic locality} (ML)~\cite{navascues2010glance}, for that the maximization of~\eqref{eq:game} can be expressed as a semi-definite program~\cite{vandenberghe1996semidefinite}, and efficiently computed; in this work we denote this set as $\mathfrak{M}$. One can show that
	\begin{equation}
		\mathfrak{L} \subset \mathfrak{S} \subset \mathfrak{Q} \subset \mathfrak{M} \subset \mathfrak{N}.
	\end{equation}
	On the other hand, one can easily see that the maximal value of~\eqref{eq:BellIneq} does not change if maximization is performed over $\mathfrak{S}$ instead of $\mathfrak{L}$.
	
	The fact that $\mathfrak{S} \subset \mathfrak{Q}$ allows for manifestation of so-called \textit{quantum non-locality}, or \textit{Bell non-locality}~\cite{brunner2014bell}. It was first observed in 1935 by Einstein, Podolsky, and Rosen~\cite{einstein1935can}, and considered as a caveat of the quantum theory, and later reconsidered by Bell in 1964~\cite{bell1964einstein}, who derived operational consequences of this phenomenon, first shown experimentally in the 1980s by Aspect~\cite{aspect1982experimental,aspect1982experimental2}, later in a loophole-free realization in series of works in 2015~\cite{hensen2015loophole,giustina2015significant}, and in a recent spectacular experiment with entangled photon distributed over satellites~\cite{yin2017satellite}.

	\section{Related works and scope}
	\label{sec:related}
	
	The problem of coordinating actions without communication but with the assistance of entangled quantum states in multi-partite games falls into the general category of quantum pseudo-telepathy~\cite{brassard1999cost,brassard2005quantum}.
	
	The closest to our work is the paper~\cite{brukner2006entanglement}, where Bell-type inequalities were used in the context of arranging the meeting of two parties that were not able to communicate. In the paper Alice and Bob were located on the North and the South Pole and had $6$ possible paths to reach the Equator, each path separated by a latitude of $60^{\circ}$. Their task was to use an entangled state to choose their paths so that the latitude difference after reaching the Equator was not greater than $60^{\circ}$. The authors assumed that the parties could not pre-agree on the chosen paths, and they both knew the starting locations of both of them. Thus this was not a rendezvous scenario, and the limit that the parties had fixed known initial locations reduces the applicability of that scenario.
	
	Quantum cooperation between animals has been hypothesized in~\cite{summhammer2005quantum,summhammer2011quantum}. In those works, insects are assumed to share multiple copies of maximally entangled states to correlate their probabilistic moves in order to find each other. The insects performed short flights with a support of a presence of a weak scent from the other insect to provide an approximate orientation of where to fly. Their decision to either wait or move was based on the result of a quantum measurement. A similar approach was applied to ants coordinating their moves when pushing a pebble~\cite{nakayama2010experimental,iimura2014experimental,iimura2015effect}.
	
	In~\cite{khoshnoud2017quantum,khoshnoud2019quantum} a quantum communication between robots, where a leading robot transmits entangled states to follower robots in order to communicate information about cooperative tasks, is considered. In~\cite{khoshnoud2017quantum} collaborating robots perform subsequent quantum measurements on multiple shared entangled states, where measurements depend on the direction of the move to be chosen, and the result is used to determine whether to perform the action, but the current location is not taken into account in the measurement, and the direction prediction is based on classical computation. In~\cite{khoshnoud2019quantum} quantum channels were used to enhance security using a variant of BB84 cryptographic protocol~\cite{bennet1984quantum} and speed of communication, but no Bell-type nonlocality was analyzed; see~\cite{khoshnoud2020quantum} for the experimental realization of that concept. In~\cite{khoshnoud2020quantum2,khoshnoud2020quantum3} quantum teleportation protocol~\cite{bennett1993teleporting} for safe communication between robots was implemented.
	
	In~\cite{kumar2021survey} an extensive survey of the current application of quantum technologies for drones is given.
	
	In this work, we consider a rendezvous problem over graphs. The two parties are asymmetric, move synchronously, are location-aware, and have step number limit.

	\section{Methods}
	\label{sec:methods}
	
	The figure of merit considered in this paper is the success probability that two parties, Alice and Bob, will meet on a known graph $\mathcal{G}$ in a given number of steps $N_\text{max}$, and their initial positions are not known \textit{a priori} and are uniformly randomized before the first step. This refers to a real-world situation when the graph models a structure of focal points in a given area, and parties pre-agreed that they wait for each other at full hours. Let $N$ denote the number of nodes in the graph.
	
	We consider this problem with the following attributes:
	\begin{enumerate}
		\item The parties may have a different maximal allowed number of steps,
		\item The parties may or may not be allowed to choose not to move,
		\item The parties may or may not be allowed to meet on edges (upon nodes transposition),
		\item The parties may or may not start at the same node.
	\end{enumerate}
	
	In the models that we consider, the two parties move synchronously, so the first attribute effectively means a time limit, so parties have to meet before the deadline.
	
	We consider both the case when a party is allowed to decide to wait (or, not to move); this is modeled by considering a reflexive graph, and by an anti-reflexive graph in the opposite case. This refers to a situation when it is physically not possible for a party to stop for a longer time, e.g. when there are no parking places, or when high inertia forces the vehicle to move continuously.
	
	Meeting on the edges reflex possibility that there is only one physical way directly connecting the pair of neighboring focal points, thus when the parties transpose these nodes, then they will meet on their way. Further, we denote the boolean variable indicating if the parties can meet on edges by $E$.
	
	The last attribute, regarding the possibility of starting in the same node, is to be understood in a way that parties do not perform checking if the other party is present at the same place in step $0$. The real-world interpretation is that parties may start with a slight time shift, so they are not present at exactly the same time $0$, or that they start in a vicinity of a focal point, so may not notice each other. Note that if the parties were allowed to start at the same place and check rendezvous immediately, then all success probabilities would simply be rescaled, and this would not differentiate different rendezvous strategies and resources. The boolean variable to indicate if the parties can be randomized to start in the same node is denoted further as $S$.
	
	If the number $N_\text{max}$ is equal to or greater than the diameter of the considered graph, and the parties are either allowed to wait or meet on edges, then the success probability is always $1$. On the other hand, if the parties do not possess these capabilities, then for some starting locations, when the distance between them is odd, it may not be possible to rendezvous, no matter how many steps are performed.
	
	We model the strategy to be used by Alice and Bob using a joint probability distribution $P(a,b|x,y)$, where $x, y \in \{1, \cdots, N\}$ and $a \in A$ and $b \in B$ for some $A$ and $B$. This joint distribution should be understood in the following way. In their time $0$ both parties input as settings their starting locations to their parts of the box, and then they proceed in a way resulting from some deterministic function of their respective outcomes.
	
	For the sake of simplicity, in this paper we consider only regular graphs of degree $R$, and $A = B = \{1, \cdots R^{N_\text{max}} \}$, and the outcome of Alice $a$ directly determines the sequence of $N_\text{max}$ numbers $(a_s)_s$, each in $\{1, \cdots, R\}$, and each $a_s$ determines towards which edge of the currently occupied node in the step $s$ to head in the next $s + 1$ step; the same for Bob.
	
	The evaluation of the success probability is equivalent to the evaluation of a game~\eqref{eq:game} for the used box. The game can be obtained using the Algorithm~\ref{alg:GetRendezvousGameFromGraphAdjacencyList}.
	
	\begin{algorithm}[H] 
		\caption{Get a rendezvous game (a Bell operator) from a graph adjacency list}
		\label{alg:GetRendezvousGameFromGraphAdjacencyList}
		\begin{algorithmic}[1]
			\Require{graph adjacency list $L$, $N_\text{max}$, boolean for allowed edge mitting $E$, boolean for allowed same setting $S$} 
			\Ensure{$\{ B_{a,b,x,y} \}$ (coefficients of a Bell-type game)}
			\Statex
			\Function{GetGame}{$L, N_\text{max}, E, S$}
				\State{$N \gets$ number of vertices in the graph}
				\State{$R \gets$ degree of vectices in the graph}
				\If{S} \Comment{average over settings}
					\State{$p \gets 1 / N^2$}
				\Else
					\State{$p \gets 1 / (N \cdot (N - 1))$}
				\EndIf
				\For{$a \gets 1$ to $R^{N_\text{max}}$}
					\For{$b \gets 1$ to $R^{N_\text{max}}$}
						\For{$x \gets 1$ to $N$}
							\If{$S$}
								\State{$\tilde{Y} \gets \{1, \cdots N\}$}
							\Else
								\State{$\tilde{Y} \gets \{1, \cdots N\} \setminus x$} 
							\EndIf
							\For {$y \in \tilde{Y}$}
								\State{$p_A \gets x$} \Comment{current positions}
								\State{$p_B \gets y$}
								\For{$s \gets 1$ to $N_\text{max}$} \Comment{steps}
									\State{$n_A \gets L(p_A, a_s)$} \Comment{$a_s$-th edge of $p_A$}
									\State{$n_B \gets L(p_B, b_s)$} \Comment{$b_s$-th edge of $p_B$}
									\If{$n_A = n_B$ or ($E$ and $p_A = n_B$ and $p_B = n_A$)} \Comment{meet or transpose}
										\State{$B_{a,b,x,y} \gets p$}
									\EndIf
									\State{$p_A \gets n_A$} \Comment{move to next positions}
									\State{$p_B \gets n_B$}
								\EndFor
							\EndFor
						\EndFor
					\EndFor
				\EndFor
				\State \Return {$\{ B_{a,b,x,y} \}$}
			\EndFunction
		\end{algorithmic}
	\end{algorithm}

	To evaluate the gain possible to be obtained when the parties are assisted with a quantum entangled state we calculated the Tsirelon's bound for Alice and Bob using different classes of resources, \textit{viz.} with boxes from sets $\mathfrak{S}$, $\mathfrak{Q}$, $\mathfrak{M}$, and $\mathfrak{N}$. To this end, we wrote MATLAB scripts with crucial parts implemented in C++ using the MEX technology.
	
	For maximization over $\mathfrak{S}$ we enumerated all possible deterministic strategies. For a single party there are $R^{N_\text{max}}$ outcomes and $N$ settings, meaning $R^{N \cdot N_\text{max}}$ deterministic functions $\{1, \cdots, N\} \to \{1, \cdots, R^{N_\text{max}}\}$. Thus, for two asymmetric parties, we needed to consider the game value for $R^{2 N \cdot N_\text{max}}$ deterministic boxes.
	
	A lower bound for what can be obtained using quantum resources, we searched for explicit state $\rho$, and measurements $\{ \{ M^a _x\}_a \}_x$, and $\{ \{ N^b _y\}_b \}_y$ on fixed Hilbert spaces $\mathcal{H}^{(A)}$ and $\mathcal{H}^{(B)}$. We employed the see-saw optimization technique~\cite{pal2010maximal}, where one intertwine linear optimizations separately over each variable in the non-linear formulae~\eqref{eq:PabxyQ}. This provides only a lower bound, as the method is not guaranteed to reach the global optimum, yet it suffices to show the advantage in many cases. Still, one should note that the more complicated the case, \textit{i.e.} the more settings and outcomes and the larger dimension used, the smaller are chances that the method yields the maximal value.
	
	An upper bound for the gain from using the quantum resources is got from the maximization over the set $\mathfrak{M}$ of the ML correlations mentioned above. A direct numerical method to perform this optimization is the \textit{Navascu{\'e}s-Pironio-Acin}~\cite{navascues2007bounding,navascues2008convergent} semi-definite programming method. These correlations may not be possible to be realized in practical implementations, but determines the scope of what possibly could be reached.
	
	Finally, we also derived the non-signaling bound of the set $\mathfrak{N}$ using linear programming. This value has (almost~\cite{colbeck2009quantum}) no practical application, but provides an insight regarding the meaning of quantum limits on the rendezvous tasks.
	
	Below we use the following notation for box outcomes. We assume the edges of the graphs to be labeled from $1$ to the vertex's degree in such a way that the labels of the vertices towards that relevant edges are connecting increase with the edge's number. For instance, if the vertex $3$ has edges leading to vertices $1$ and $5$, and an edge towards itself (e.g. in reflexive graphs), then these edges will have labels $1$, $3$ and $2$, respectively.

	\section{Results}
	\label{sec:results}
	
	We performed the analysis over three sets of graphs of small size: cubic (3-regular) graphs with 6 or 8 nodes, see sec.~\ref{sec:cubic} and Listing~\ref{lst:cubic}; cycles over at most 9 nodes with step number limit $1$, see sec.~\ref{sec:cycle1}, and $2$, see sec.~\ref{sec:cycle2}; and directed cycles, see sec.~\ref{sec:directed}. In the cases when the parties are allowed to choose not to move, we used reflexive versions of these graphs. In sec.~\ref{sec:robustness} we analyze the robustness and experimental feasibility of our results.
	
	\subsection{Cubic graphs with single step}
	\label{sec:cubic}
	
	\begin{lstlisting}[caption={Adjacency lists of cubic graphs with 6 or 8 nodes},label={lst:cubic},float,captionpos=b]
cubic-2 = [2 3 4; 1 3 5; 1 2 6; 1 5 6;
			2 4 6; 3 4 5]
cubic-3 = [4 5 6; 4 5 6; 4 5 6; 1 2 3;
			1 2 3; 1 2 3]
cubic-4 = [3 5 7; 4 6 8; 1 5 7; 2 6 8;
		1 3 7; 2 4 8; 1 3 5; 2 4 6]
cubic-5 = [2 5 6; 1 3 6; 2 4 7; 3 5 8;
		1 4 8; 1 2 7; 3 6 8; 4 5 7]
cubic-6 = [2 3 4; 1 3 6; 1 2 8; 1 5 7;
		4 6 8; 2 5 7; 4 6 8; 3 5 7]
cubic-7 = [2 4 5; 1 3 6; 2 4 7; 1 3 8;
		1 6 8; 2 5 7; 3 6 8; 4 5 7]
cubic-8 = [2 6 7; 1 3 7; 2 4 7; 3 5 8;
		4 6 8; 1 5 8; 1 2 3; 4 5 6]
cubic-9 = [2 5 8; 1 3 6; 2 4 7; 3 5 8;
		1 4 6; 2 5 7; 3 6 8; 1 4 7]
	\end{lstlisting}

	Let us consider the case with single step restriction, \textit{i.e.} $N_\text{max} = 1$. For cubic graphs, we consider only a single step case due to numerical complexity for cases with more vertices.
	
	For the 6 node cubic graphs, labeled in Listing~\ref{lst:cubic} as 2 and 3, we observed no gain if the parties were not allowed to meet, and some gain in the opposite case when quantum resources were used. The gains for ML and NS correlation were, on contrary, very strong, up to $28.6\%$. A similar gain in the case with allowed waiting was present also for cubic graphs labeled 6 and 9, and here the gain with quantum resources was up to $9.2\%$, and with NS was up to $33.3\%$; see Tab.~\ref{lab:cubic2369reflexive}. The cubic graphs 6 and 9 showed also quantum gain when the waiting was not allowed.
	
	\begin{table}[]
		\begin{tabular}{|l|l|l|l|l|l|l|}
			\hline
			Set & $E$ & $S$ & cubic-2 & cubic-3 & cubic-6 & cubic-9 \\ \hline
			$\mathfrak{S}$ & 0 & 0 & 0.46667 & 0.46667 & 0.32143 & 0.32143 \\
			$\mathfrak{Q}$ & 0 & 0 & 0.46676 & 0.46676 & 0.33656 & 0.35101 \\
			$\mathfrak{M}$ & 0 & 0 & 0.50014 & 0.50014 & 0.3587 & 0.3585 \\
			$\mathfrak{N}$ & 0 & 0 & 0.6 & 0.6 & 0.42857 & 0.42857 \\ \hline
			$\mathfrak{S}$ & 1 & 0 & 0.46667 & 0.46667 & 0.35714 & 0.32143 \\
			$\mathfrak{Q}$ & 1 & 0 & 0.47072 & 0.46978 & fail & 0.35101 \\
			$\mathfrak{M}$ & 1 & 0 & 0.50356 & 0.51287 & 0.36268 & 0.36108 \\
			$\mathfrak{N}$ & 1 & 0 & 0.6 & 0.6 & 0.42857 & 0.42857 \\ \hline
			$\mathfrak{S}$ & 0 & 1 & 0.55556 & 0.55556 & 0.40625 & 0.40625 \\
			$\mathfrak{Q}$ & 0 & 1 & 0.55564 & 0.55564 & 0.41831 & 0.43214 \\
			$\mathfrak{M}$ & 0 & 1 & 0.57579 & 0.57579 & 0.43625 & 0.43651 \\
			$\mathfrak{N}$ & 0 & 1 & 0.66667 & 0.66667 & 0.5 & 0.5 \\ \hline
			$\mathfrak{S}$ & 1 & 1 & 0.55556 & 0.55556 & 0.40625 & 0.40625 \\
			$\mathfrak{Q}$ & 1 & 1 & 0.55857 & 0.55819 & 0.41726 & 0.43214 \\
			$\mathfrak{M}$ & 1 & 1 & 0.57743 & 0.58352 & 0.43712 & 0.43665 \\
			$\mathfrak{N}$ & 1 & 1 & 0.66667 & 0.66667 & 0.5 & 0.5 \\ \hline
		\end{tabular}
		\caption{Comparison of efficiency of sets $\mathfrak{S}$, $\mathfrak{Q}$, $\mathfrak{M}$, and $\mathfrak{N}$, for selected cubic graphs when parties are allowed to wait and $N_\text{max} = 1$. In one case the see-saw method failed to find a reliable result.}
		\label{lab:cubic2369reflexive}
	\end{table}
	
	In contrast to 6 node cubic graphs, for $N_\text{max} = 1$ all 8 node cubic graphs showed an advantage when the parties were not allowed to wait, see Tab.~\ref{tab:cubic456789antireflexive}. The gains are largest both in quantum and NS cases when parties were not possible to be randomized to start at the same position and were allowed to meet on edges ($S = 0$ and $E = 1$); the gains were smallest in the opposite case. The quantum gain varied between tiny $0.2\%$ and $16.7\%$, whereas the NS gain was between $9.1\%$ and $50\%$.
	
	\begin{table}[]
		\begin{tabular}{|l|l|l|l|l|l|l|l|l|}
			\hline
			Set            & $E$ & $S$ & cubic-4 & cubic-5 & cubic-6 & cubic-7 & cubic-8 & cubic-9 \\ \hline
			$\mathfrak{S}$ & 0   & 0   & 0.21429 & 0.25    & 0.25    & 0.21429 & 0.21429 & 0.25    \\
			$\mathfrak{Q}$ & 0   & 0   & 0.22857 & fail   & 0.25303 & 0.22857 & 0.22857 & 0.26546 \\
			$\mathfrak{M}$ & 0   & 0   & 0.2381  & 0.25462 & 0.25893 & 0.2381  & 0.24478 & 0.26749 \\
			$\mathfrak{N}$ & 0   & 0   & 0.28571 & 0.28571 & 0.28571 & 0.28571 & 0.28571 & 0.28571 \\ \hline
			$\mathfrak{S}$ & 1   & 0   & 0.28571 & 0.28571 & 0.28571 & 0.28571 & 0.28571 & 0.28571 \\
			$\mathfrak{Q}$ & 1   & 0   & 0.33333 & 0.32087 & 0.31338 & 0.33333 & 0.33333 & 0.30764 \\
			$\mathfrak{M}$ & 1   & 0   & 0.33333 & 0.33063 & 0.32651 & 0.33333 & 0.33333 & 0.32951 \\
			$\mathfrak{N}$ & 1   & 0   & 0.42857 & 0.42857 & 0.42857 & 0.42857 & 0.42857 & 0.42857 \\ \hline
			$\mathfrak{S}$ & 0   & 1   & 0.3125  & 0.34375 & 0.34375 & 0.3125  & 0.3125  & 0.34375 \\
			$\mathfrak{Q}$ & 0   & 1   & 0.32253 & fail   & 0.34604 & 0.32253 & 0.32252 & 0.35728 \\
			$\mathfrak{M}$ & 0   & 1   & 0.325   & 0.34734 & 0.35156 & 0.325   & 0.33098 & 0.35898 \\
			$\mathfrak{N}$ & 0   & 1   & 0.375   & 0.375   & 0.375   & 0.375   & 0.375   & 0.375   \\ \hline
			$\mathfrak{S}$ & 1   & 1   & 0.375   & 0.375   & 0.375   & 0.375   & 0.375   & 0.375   \\
			$\mathfrak{Q}$ & 1   & 1   & 0.39815 & 0.38884 & 0.37568 & 0.39815 & 0.39815 & 0.38299 \\
			$\mathfrak{M}$ & 1   & 1   & 0.4     & 0.39679 & 0.38332 & 0.4     & 0.4     & 0.38535 \\
			$\mathfrak{N}$ & 1   & 1   & 0.5     & 0.5     & 0.5     & 0.5     & 0.5     & 0.5     \\ \hline
		\end{tabular}
		\caption{Comparison of efficiency of sets $\mathfrak{S}$, $\mathfrak{Q}$, $\mathfrak{M}$, and $\mathfrak{N}$, for cubic graphs with 8 nodes when parties are not allowed to wait and $N_\text{max} = 1$. In two cases the see-saw method failed to find a reliable result.}
		\label{tab:cubic456789antireflexive}
	\end{table}

	\subsection{Cycle graphs with single step}
	\label{sec:cycle1}

	The cycle graphs when $N_\text{max} = 1$, there was no gain in most of the cases when $S = 0$, with exception for the cycle over $4$, $5$, $7$, and $8$ nodes with the possibility of waiting and without edge meeting, see Tab.~\ref{tab:cycle4578_S0}.
	
	Let us take a look at strategies optimal for the cycle over $4$ nodes in this scenario. It reveals that the optimal strategy is for both parties to take the edge that leads to the node with smallest number, \textit{i.e.} to move from nodes $1$, $2$, $3$, and $4$ towards nodes $1$ (wait), $1$, $2$ and $1$, respectively, \textit{cf.}~\cite{howard1999rendezvous}. Direct calculation shows that they succeed in $50\%$, \textit{viz.} whenever their starting positions were given by pairs $(1, 2)$, $(1, 4)$, $(2, 1)$, $(2, 4)$, $(4, 1)$, or $(4, 2)$, and that they fail if these were $(1, 3)$, $(2, 3)$, $(3, 1)$, $(3, 2)$, $(3, 4)$, or $(4, 3)$. The quantum box is tedious to be written explicitly, and the NS is given by
	\begin{subequations}
		\label{eq:NSbox_cycle4_S0}
		\begin{equation}
			\forall_{a,b,x} P(a,b,x,x) = 1/9,
		\end{equation}
		\begin{equation}
			\begin{aligned}
				\forall_{a} P&(a,a,2,1) = P(a,a,4,1) = P(a,a,1,2) = \\
				&P(a,a,4,3) = P(a,a,1,4) = P(a,a,3,4) = 1/3,
			\end{aligned}
		\end{equation}
		\begin{equation}
			\forall_{a,b} P(a,b,3,1) = P(a,b,1,3) = B_1(a,b),
		\end{equation}
		\begin{equation}
			\forall_{a,b} P(a,b,2,4) = P(a,b,4,2) = B_2(a,b),
		\end{equation}
		\begin{equation}
			\begin{aligned}
				P&(3,1,3,2) = P(1,2,3,2) = P(2,3,3,2) = \\
				&P(2,1,2,3) = P(3,2,2,3) = P(1,3,2,3) = 1/3,
			\end{aligned}
		\end{equation}
	\end{subequations}
	where $B_1(2,1) = B_1(1,2) = B_1(3,3) = 1/3$ and $B_2(1,1) = B_2(3,2) = B_2(2,3) = 1/3$, and all remaining values are equal $0$.
	
	\begin{table}[]
		\begin{tabular}{|l|l|l|l|l|}
			\hline
			Set            & Cycle 4 & Cycle 5 & Cycle 7 & Cycle 8 \\ \hline
			$\mathfrak{S}$ & 0.5     & 0.4     & 0.28571 & 0.25    \\
			$\mathfrak{Q}$ & 0.53333 & 0.4117  & 0.30709 & 0.26546 \\
			$\mathfrak{M}$ & 0.55556 & 0.42888 & 0.30896 & 0.26748 \\
			$\mathfrak{N}$ & 0.66667 & 0.5     & 0.33333 & 0.28571 \\ \hline
		\end{tabular}
		\caption{Comparison of efficiency of sets $\mathfrak{S}$, $\mathfrak{Q}$, $\mathfrak{M}$, and $\mathfrak{N}$, for cycle graphs with $4$, $5$, $7$, and $8$ nodes, when the parties are allowed to wait, no edge meeting is allowed ($E = 0$), parties are always randomized at different initial positions ($S = 0$), and $N_\text{max} = 1$.}
		\label{tab:cycle4578_S0}
	\end{table}
	
	Now, we consider the case when the parties may be initially located at the same initial position, \textit{i.e.} in a vicinity of the same focal point, $S = 1$. In this case, we observed a gain in the majority of the cases with cycle graphs.
	
	The results for the case with anti-reflexive graphs, \textit{i.e.} when parties are not allowed to wait, are shown in Tab.~\ref{tab:cycle_S1}. For cycles of length $4$ and $8$, there was no quantum gain, and for the other case, the quantum gain varies between $2.3\%$ and $7.1\%$, whereas NS gains are within the range from $5.9\%$ to $36.4\%$.
	
	\begin{table}[]
		\scriptsize
		\begin{tabular}{|l|l|l|l|l|l|l|l|l|}
			\hline
			Set            & $E$ & Cycle 3 & Cycle 4 & Cycle 5 & Cycle 6 & Cycle 7 & Cycle 8 & Cycle 9 \\ \hline
			$\mathfrak{S}$ & 0   & 0.55556 & 0.5     & 0.36    & 0.27778 & 0.26531 & 0.25    & 0.20988 \\
			$\mathfrak{Q}$ & 0   & 0.58333 & 0.5     & 0.3809  & 0.29167 & 0.27864 & 0.25    & 0.21887 \\
			$\mathfrak{M}$ & 0   & 0.58333 & 0.5     & 0.3809  & 0.29167 & 0.27864 & 0.25    & 0.21887 \\
			$\mathfrak{N}$ & 0   & 0.66667 & 0.5     & 0.4     & 0.33333 & 0.28571 & 0.25    & 0.22222 \\ \hline
			$\mathfrak{S}$ & 1   & 0.77778 & 0.625   & 0.44    & 0.38889 & 0.34694 & 0.3125  & 0.25926 \\
			$\mathfrak{Q}$ & 1   & 0.83333 & 0.625   & 0.45    & 0.41667 & 0.36596 & 0.3125  & 0.27778 \\
			$\mathfrak{M}$ & 1   & 0.83333 & 0.625   & 0.45    & 0.41667 & 0.36596 & 0.3125  & 0.27778 \\
			$\mathfrak{N}$ & 1   & 1       & 0.75    & 0.6     & 0.5     & 0.42857 & 0.375   & 0.33333 \\ \hline
		\end{tabular}
		\caption{Comparison of efficiency of sets $\mathfrak{S}$, $\mathfrak{Q}$, $\mathfrak{M}$, and $\mathfrak{N}$, for cycle graphs when parties are not allowed to wait, and are possibly randomized at the same initial positions ($S = 1$), and $N_\text{max} = 1$.}
		\label{tab:cycle_S1}
	\end{table}

	The cases of length $4$ and $8$ are interesting, as the NS advantage is possible there even though quantum and ML sets yield the same value as LHV. Similarly, as in the case $S = 0$ described above, the optimal LHV is for both parties to move toward the nodes with a smaller number; anti-reflexivity implies that a party from node $1$ should move to node $2$. For the NS set the optimal strategy, \textit{cf.}~\eqref{eq:NSbox_cycle4_S0}, is to use a box with $P(a, \neg a|x,y) = 0.5$ for $(x,y) \in \{(1,4), (4,1), (2,3), (3,2)\}$, and with $P(a,a|x,y) = 0.5$ for other pairs of settings.
	
	For the case when the parties are allowed to wait and are possibly starting at the same position, the results are shown in Tab.~\ref{tab:cycle_S1_reflexive}, and the gain is observed, similarly as for the case when the parties always start at a different position, \textit{cf.} Tab.~\ref{tab:cycle4578_S0}, only for cycles over $4$, $5$, $7$ and $9$ nodes. The quantum gain is between $1.8\%$ and $4.8\%$, and the NS gain is between $9.1\%$ and $20\%$. Note that for the cycle over $4$ nodes and $E = 1$ the LHV and NS values are the same no matter if the parties are not allowed to wait or not, whereas quantum and ML gains become possible in this case when the parties are allowed to wait; this means that the quantum advantage comes from a possibility to better coordinate which party should be waiting.

	\begin{table}[]
		\begin{tabular}{|l|l|l|l|l|l|}
			\hline
			Set            & $E$ & Cycle 4 & Cycle 5 & Cycle 7 & Cycle 8 \\ \hline
			$\mathfrak{S}$ & 0   & 0.625   & 0.52    & 0.38776 & 0.34375 \\
			$\mathfrak{Q}$ & 0   & 0.64506 & 0.52936 & 0.40607 & 0.35728 \\
			$\mathfrak{M}$ & 0   & 0.65    & 0.54007 & 0.40719 & 0.35898 \\
			$\mathfrak{N}$ & 0   & 0.75    & 0.6     & 0.42857 & 0.375   \\ \hline
			$\mathfrak{S}$ & 1   & 0.625   & 0.52    & 0.38776 & 0.34375 \\
			$\mathfrak{Q}$ & 1   & 0.64872 & 0.53129 & 0.40631 & 0.35745 \\
			$\mathfrak{M}$ & 1   & 0.65491 & 0.54016 & 0.40774 & 0.3591  \\
			$\mathfrak{N}$ & 1   & 0.75    & 0.6     & 0.42857 & 0.375 \\ \hline 
		\end{tabular}
		\caption{Comparison of efficiency of sets $\mathfrak{S}$, $\mathfrak{Q}$, $\mathfrak{M}$, and $\mathfrak{N}$, for cycle graphs $4$, $5$, $7$, and $8$, when parties are allowed to wait, and are possibly randomized at the same initial positions ($S = 1$), and $N_\text{max} = 1$.}
		\label{tab:cycle_S1_reflexive}
	\end{table}

	\subsection{Cycle graphs with two steps}
	\label{sec:cycle2}
	
	For the cycle over $4$ vertices, when the time limit of the parties is two, $N_\text{max} = 2$, the parties can trivially meet simply by pre-agreeing about the focal point (if waiting is allowed). In general, this is true for any graphs having a radius less or equal to $N_\text{max}$. On the other hand, the complexity of calculating the capabilities of boxes from each set $\mathfrak{S}$, $\mathfrak{Q}$, $\mathfrak{M}$, and $\mathfrak{N}$ grows exponentially with $N_\text{max}$, with base $R$ (the degree of nodes in the graph) as noted in sec.~\ref{sec:methods}. For these reasons we consider only the case when the parties are not allowed to wait (\textit{i.e.} anti-reflexive graphs), and cycle graphs of sizes $5$, $6$, $7$, and $8$.
	
	The calculations revealed that a gain is obtained only when the parties can be randomized to the same starting location, $S = 1$. The results are shown in Tab.~\ref{tab:cycle_Nmax2_S1_antireflexive}.
	
	\begin{table}[]
		\begin{tabular}{|l|l|l|l|l|l|}
			\hline
			Set            & $E$ & Cycle 5 & Cycle 6 & Cycle 7 & Cycle 8 \\ \hline
			$\mathfrak{S}$ & 0   & 0.52    & 0.5     & 0.38776 & 0.3125  \\
			$\mathfrak{Q}$ & 0   & 0.52234 & 0.5     & 0.38776 & 0.3125  \\
			$\mathfrak{M}$ & 0   & 0.55013 & 0.5     & 0.41273 & 0.34506 \\
			$\mathfrak{N}$ & 0   & 0.6     & 0.5     & 0.42857 & 0.375   \\ \hline
			$\mathfrak{S}$ & 1   & 0.84    & 0.72222 & 0.59184 & 0.5     \\
			$\mathfrak{Q}$ & 1   & 0.89271 & 0.72222 & 0.59184 & 0.5     \\
			$\mathfrak{M}$ & 1   & 0.90076 & 0.75    & 0.62478 & 0.53178 \\
			$\mathfrak{N}$ & 1   & 1       & 0.83333 & 0.71429 & 0.625  \\ \hline
		\end{tabular}
		\caption{Comparison of efficiency of sets $\mathfrak{S}$, $\mathfrak{Q}$, $\mathfrak{M}$, and $\mathfrak{N}$, for cycle graphs $5$, $6$, $7$, and $8$, when parties are not allowed to wait (anti-reflexive graphs), and are possibly randomized at the same initial positions ($S = 1$), and $N_\text{max} = 2$.}
		\label{tab:cycle_Nmax2_S1_antireflexive}
	\end{table}
	
	Interestingly, for cycles over $5$, $6$, and $7$ nodes the value obtained with see-saw for the quantum set $\mathfrak{Q}$ is exactly equal to the LHV value, whereas the ML and NS values are greater than LHV. The result has been confirmed by subsequent reevaluations of the see-saw with different initializations, but still provides only a lower bound, thus the actual quantum capability can be larger.
	
	A quantum gain is possible for the $5$ node cycle both in cases with edge meeting allowed, $E = 1$, and not, $E = 0$. We note that the success probability in NS with $E = 1$ is $1$, meaning that there exists a non-signaling box that always shows to the parties the proper direction so that they approach each other. Below we analyze this case in more detail.
	
	For Alice starting in the node $1$ and Bob starting in the node $5$, an NS box assures that:
	\begin{enumerate}
		\item If Bob visits the node $1$ and then the node $2$, then Alice visits either nodes $2$ and then $1$, or $5$ and then $1$ or $5$ and then $4$; thus they will either meet transposing nodes $1$ and $2$ in the first case, or $1$ and $5$ in the second and third case.
		\item If Bob visits the node $1$ and then the node $5$, then Alice visits either node $5$ and then $1$, or $5$ and then $4$; in both cases, they will meet transposing nodes $1$ and $5$.
		\item If Bob visits node $4$ and then the node $3$, then Alice visits nodes $2$ and then $3$; they will meet after two steps at node $3$.
		\item If Bob visits the node $4$ and then the node $5$, then Alice visits nodes $5$ and $4$; they will meet after two steps transposing nodes $4$ and $5$.
	\end{enumerate}
	For LHV boxes, an optimal strategy is to always choose the node with a lower labeling number. For the initial positions $1$ and $5$ this behaves as the first option for the described NS box, \textit{i.e.} Bob visits nodes $1$ and $2$, and Alice visits nodes $2$ and $1$, and they meet on transposing nodes $1$ and $2$. For the quantum box, we have two possibilities for Bob, and for both Alice can react in two ways:
	\begin{itemize}
		\item When Bob visits the node $1$ and then the node $5$, then Alice with \textit{low} probability visits $2$ and then $3$ or with \textit{high} probability visits $5$ and then $1$; thus with \textit{high} probability they meet transposing nodes $1$ and $5$ after one step.
		\item When Bob visits the node $4$ and then the node $3$, then Alice with \textit{high} probability visits $2$ and then $3$ or with \textit{low} probability visits $5$ and then $1$; thus with \textit{high} probability they meet at the node $3$ after two steps.
	\end{itemize}
	
	Whereas for the pair of settings $1$ and $5$ both LHV and NS strategies win with certainty, and the quantum strategy wins with high probability, the LHV fails \textit{e.g.} for the pair of settings $1$ and $4$. On contrary, for these settings, the NS box satisfies:
	\begin{enumerate}
		\item If Bob visits the node $3$ and then the node $2$, then Alice visits nodes $2$ and then $3$; the parties will meet transposing nodes $2$ and $3$.
		\item If Bob visits node $3$ and then the node $4$, then Alice visits nodes $5$ and then $4$; the parties will meet at node $4$.
		\item If Bob visits the node $5$ and then the node $1$, then Alice visits either nodes $2$ and then $1$, or $5$ and then $1$, or $5$ and then $4$; they will meet either after two steps at the node $1$ in the first case or after one step at the node $5$ in the second and third case.
		\item If Bob visits the node $5$ and then the node $4$, then Alice visits either node $5$ and then $1$, or $5$ and then $4$; in both cases, they meet in the first step at the node $5$.
	\end{enumerate}
	This ensures that the NS box wins with a probability of $1$. For the quantum box, similarly, as in the previous case we have the following possibilities:
	\begin{itemize}
		\item When Bob visits the node $3$ and then the node $2$, then Alice with \textit{high} probability visits $2$ and then $3$ or with \textit{low} probability visits $5$ and then $1$; thus with \textit{high} probability they meet transposing nodes $2$ and $3$ after two steps.
		\item When Bob visits the node $5$ and then the node $4$, then Alice with \textit{low} probability visits $2$ and then $3$ or with \textit{high} probability visits $5$ and then $1$; thus with \textit{high} probability, they meet at the node $5$ after one step.
	\end{itemize}
	The quantum box that we consider provides success for any pair of settings with probabilities between $0.83$ and $1$. We observe that the probability $1$ is obtained exactly for these settings where $y = x + 1$, with $6 \equiv 1$, \textit{i.e.} for $(x,y) \in \{(1,2), (2,3), (3,4), (4,5), (5,1)\}$.

	\subsection{Directed cycles with two steps}
	\label{sec:directed}
	
	As the last example we consider reflexive directed cycle graphs and two-step limit, $N_\text{max} = 2$. The reflexive directed cycle graphs over $N$ nodes has edges of the form $(i,i)$ and $(i,i+1)$, where $i \in \{1, \cdots, N\}$, and $N + 1 \equiv 1$. This means that from each node a party can either wait or move in the node increasing number direction.
	
	By using these graphs we reduce the complexity of the problems to be analyzed, \textit{viz.} for $N_\text{max} = 2$ a box has to output one of only four possible outcomes, that are further interpreted as instructions for two subsequent moves of the party. The formalism remains exactly the same as in the previous sections.
	
	We note that in this case there is no possibility for edge meeting, so cases with $E = 0$ and $E = 1$ are trivially equivalent. Recall that in the ordinary, bidirectional, cycles, in most of the cases with $S = 0$ we observed no quantum advantage. In the directed cycles we observed that all cases with gain has $S = 1$, see Tab.~\ref{tab:directed_cycles_steps2}. For directed cycle with $9$ nodes there was no quantum advantage, and the success probability was $1/4$ for $S = 0$, and $1/3$ for $S = 1$ for all sets $\mathfrak{S}$, $\mathfrak{Q}$, $\mathfrak{M}$, and $\mathfrak{N}$.
	
	\begin{table}[]
		\scriptsize
		\begin{tabular}{|l|l|l|l|l|l|}
			\hline
			Set            & Dir. cycle 4 & Dir. cycle 5 & Dir. cycle 6 & Dir. cycle 7 & Dir. cycle 8 \\ \hline
			$\mathfrak{S}$ & 0.625        & 0.52         & 0.5          & 0.38776      & 0.34375      \\
			$\mathfrak{Q}$ & 0.67678      & 0.52         & 0.5          & 0.39044      & 0.34717      \\
			$\mathfrak{M}$ & 0.69012      & 0.55013      & 0.5          & 0.41273      & 0.3614       \\
			$\mathfrak{N}$ & 0.75         & 0.6          & 0.5          & 0.42857      & 0.375       \\ \hline
		\end{tabular}
		\caption{Success probabilities when parties are located at random places with $S = 1$ (allowed the same starting node for both parties) of a reflexive directed cycle graph for two step limit $N_\text{max} = 2$.}
		\label{tab:directed_cycles_steps2}
	\end{table}
	
	Let us first analyze the peculiar case of a directed $6$-node cycle. An optimal LHV strategy is for instance:
	\begin{enumerate}
		\item If the party is in the node $1$ then waits for two steps.
		\item If the party is in the node $2$, then move to the node $3$ and then $4$.
		\item If the party is in the node $3$, then wait, and in the second step move to the node $4$.
		\item If the party is in the node $4$ then waits for two steps.
		\item If the party is in the node $5$, then move to the node $6$ and then $1$.
		\item If the party is in the node $6$, then move to the node $1$ and wait.
	\end{enumerate}
	This strategy is the same for both parties, Alice and Bob. Direct calculations show that this gives a success probability of $0.5$. This result cannot be improved by using quantum, ML, or NS boxes. The successful pairs of settings are $\{(1,5), (1,6), (2,3), (2,4), (3,4), (5,6)\}$, and their reverses, and all pairs $(a,a)$.
	
	An optimal LHV strategy for $4$ nodes is the following. For any setting, Alice waits at her initial node, as in the well know approach called \textit{wait-for-mummy}~\cite{howard1999rendezvous}. Bob moves in the following way:
	\begin{enumerate}
		\item If Bob is in the node $1$, then wait for one step and then move to the node $2$.
		\item If Bob is in the node $2$, then wait for one step and then move to the node $3$.
		\item If Bob is in the node $3$, then move to the node $4$ and then $1$.
		\item If Bob is in the node $4$, then move to the node $1$ and then $2$.
	\end{enumerate}
	This deterministic asymmetric strategy provides a success probability of $0.625$. If we fix Alice's part of the box to the wait-for-mummy strategy, then quantum boxes are not able to provide any advantage, whereas without this constraint there exists a quantum box that gives a success probability of at least $0.67678$.
	
	\subsection{Robustness to noise}
	\label{sec:robustness}
	
	Any future experimental realization of the proposed rendezvous method will need to be robust against noise and quantum state imperfections. To estimate what is the critical noise tolerance that still allows the proposed protocols to show quantum advantage we model the noise as an effective replacement of the maximally entangled state $\ket{\Phi}$ with the noised one given by the formula
	\begin{equation}
		\nu \proj{\Phi} + (1 - \nu) \rho_\text{mixed},
	\end{equation}
	where $\rho_\text{mixed}$ is the maximally mixed state modeling uniform classical probability distribution of all possible states, \textit{i.e.} the white noise. The critical value of $\nu$ denoted $\nu_\text{crit}$ is defined as the level of noise when their quantum strategy has the same efficiency as the LHV strategy.
	
	To illustrate the robustness we considered the case of cubic graphs from Tab.~\ref{tab:cubic456789antireflexive}, \textit{viz.} cubic graphs number $4$ to $9$, anti-reflexive. For conciseness, we set $E = 1$ and $S = 0$. The values of $\nu_\text{crit}$ are $0.75$, $0.80252$, $0.83777$, $0.75$, $0.75$, $0.86695$, respectively. We obtain similar values for other cases with quantum advantage. This result shows that the proposed protocols are feasible for experimental and practical realizations using current technologies.

	\section{Conclusions}
	\label{sec:conclusions}
	
	In this paper, we have proposed a novel quantum protocol that allowed to complete the rendezvous task on a network for asymmetric, location-aware agents, and time limits with higher probability than using only classical resources. Our study covered cases of cubic graphs and cycles with $1$ and $2$ step limits. Due to the high complexity of the problem, we were not able to analyze more complicated instances and leave more optimized numerical or analytical approaches for future research.
	
	It is interesting to investigate whether a similar approach with quantum non-locality can be applied to closely related problems, like other search games~\cite{ahlswede1987search,baston2019search}, possibly with the accompanying of another quantum phenomena of the so-called Elitzur–Vaidman bomb testing or interaction free measurement~\cite{elitzur1993quantum,kwiat1995interaction}.
	
	The aim of this work was to bridge the two, till now independent, research areas of rendezvous in classical computer science, and Bell non-locality in quantum information science.

	\section*{Acknowledgments}
	The work is supported by the Foundation for Polish Science (IRAP project, ICTQT, contract no. 2018/MAB/5, co-financed by EU within Smart Growth Operational Programme), NCBiR QUANTERA/2/2020 (www.quantera.eu) under the project eDICT, and NCN grant SHENG (contract No. 2018/30/Q/ST2/00625). The numerical calculations we conducted using OCTAVE~6.1~\cite{OCTAVE}, and packages SeDuMi 1.3~\cite{sturm1999using}, SDPT3~\cite{toh1999sdpt3} and YALMIP~\cite{lofberg2004yalmip}. This work is dedicated to my Wife with wishes for many more entangled rendezvous.


\begin{thebibliography}{100}

		\bibitem{alpern2002rendezvous}
		S.~Alpern, ``Rendezvous search: A personal perspective,'' {\em Operations
		  Research}, vol.~50, no.~5, pp.~772--795, 2002.

		\bibitem{alpern1995rendezvous}
		S.~Alpern, ``The rendezvous search problem,'' {\em SIAM Journal on Control and
		  Optimization}, vol.~33, no.~3, pp.~673--683, 1995.

		\bibitem{cao2012overview}
		Y.~Cao, W.~Yu, W.~Ren, and G.~Chen, ``An overview of recent progress in the
		  study of distributed multi-agent coordination,'' {\em IEEE Transactions on
		  Industrial informatics}, vol.~9, no.~1, pp.~427--438, 2012.

		\bibitem{wang2016multi}
		X.~Wang, Z.~Zeng, and Y.~Cong, ``Multi-agent distributed coordination control:
		  Developments and directions via graph viewpoint,'' {\em Neurocomputing},
		  vol.~199, pp.~204--218, 2016.

		\bibitem{yang2021overview}
		R.~Yang, L.~Liu, and G.~Feng, ``An overview of recent advances in distributed
		  coordination of multi-agent systems,'' {\em Unmanned Systems}, pp.~1--19,
		  2021.

		\bibitem{bennett1992communication}
		C.~H. Bennett and S.~J. Wiesner, ``Communication via one-and two-particle
		  operators on einstein-podolsky-rosen states,'' {\em Physical review letters},
		  vol.~69, no.~20, p.~2881, 1992.

		\bibitem{shor1994algorithms}
		P.~W. Shor, ``Algorithms for quantum computation: discrete logarithms and
		  factoring,'' in {\em Proceedings 35th annual symposium on foundations of
		  computer science}, pp.~124--134, Ieee, 1994.

		\bibitem{grover1996fast}
		L.~K. Grover, ``A fast quantum mechanical algorithm for database search,'' in
		  {\em Proceedings of the twenty-eighth annual ACM symposium on Theory of
		  computing}, pp.~212--219, 1996.

		\bibitem{buhrman2016quantum}
		H.~Buhrman, {\L}.~Czekaj, A.~Grudka, M.~Horodecki, P.~Horodecki, M.~Markiewicz,
		  F.~Speelman, and S.~Strelchuk, ``Quantum communication complexity advantage
		  implies violation of a bell inequality,'' {\em Proceedings of the National
		  Academy of Sciences}, vol.~113, no.~12, pp.~3191--3196, 2016.

		\bibitem{preskill2018quantum}
		J.~Preskill, ``Quantum computing in the nisq era and beyond,'' {\em Quantum},
		  vol.~2, p.~79, 2018.

		\bibitem{zhong2020quantum}
		H.-S. Zhong, H.~Wang, Y.-H. Deng, M.-C. Chen, L.-C. Peng, Y.-H. Luo, J.~Qin,
		  D.~Wu, X.~Ding, Y.~Hu, {\em et~al.}, ``Quantum computational advantage using
		  photons,'' {\em Science}, vol.~370, no.~6523, pp.~1460--1463, 2020.

		\bibitem{preskill2012quantum}
		J.~Preskill, ``Quantum computing and the entanglement frontier,'' {\em arXiv
		  preprint arXiv:1203.5813}, 2012.

		\bibitem{horodecki2009quantum}
		R.~Horodecki, P.~Horodecki, M.~Horodecki, and K.~Horodecki, ``Quantum
		  entanglement,'' {\em Reviews of modern physics}, vol.~81, no.~2, p.~865,
		  2009.

		\bibitem{bell1964einstein}
		J.~S. Bell, ``On the einstein podolsky rosen paradox,'' {\em Physics Physique
		  Fizika}, vol.~1, no.~3, p.~195, 1964.

		\bibitem{koopman1946search}
		B.~O. Koopman, ``Search and screening, operations evaluation group report 56,''
		  {\em Center for Naval Analysis, Alexandria, Virginia}, 1946.

		\bibitem{alpern2006theory}
		S.~Alpern and S.~Gal, {\em The theory of search games and rendezvous}, vol.~55.
		\newblock Springer Science \& Business Media, 2006.

		\bibitem{alpern2013search}
		S.~Alpern, R.~Fokkink, L.~Gasieniec, R.~Lindelauf, and V.~Subrahmanian, {\em
		  Search theory}.
		\newblock Springer, 2013.

		\bibitem{chang2015multichannel}
		C.-S. Chang, W.~Liao, and C.-M. Lien, ``On the multichannel rendezvous problem:
		  Fundamental limits, optimal hopping sequences, and bounded
		  time-to-rendezvous,'' {\em Mathematics of Operations Research}, vol.~40,
		  no.~1, pp.~1--23, 2015.

		\bibitem{mathew2013graph}
		N.~Mathew, S.~L. Smith, and S.~L. Waslander, ``A graph-based approach to
		  multi-robot rendezvous for recharging in persistent tasks,'' in {\em 2013
		  IEEE International Conference on Robotics and Automation}, pp.~3497--3502,
		  IEEE, 2013.

		\bibitem{otto2018optimization}
		A.~Otto, N.~Agatz, J.~Campbell, B.~Golden, and E.~Pesch, ``Optimization
		  approaches for civil applications of unmanned aerial vehicles (uavs) or
		  aerial drones: A survey,'' {\em Networks}, vol.~72, no.~4, pp.~411--458,
		  2018.

		\bibitem{leone2022search}
		P.~Leone, J.~Buwaya, and S.~Alpern, ``Search-and-rescue rendezvous,'' {\em
		  European Journal of Operational Research}, vol.~297, no.~2, pp.~579--591,
		  2022.

		\bibitem{mizumoto2019adaptive}
		N.~Mizumoto and S.~Dobata, ``Adaptive switch to sexually dimorphic movements by
		  partner-seeking termites,'' {\em Science advances}, vol.~5, no.~6,
		  p.~eaau6108, 2019.

		\bibitem{isaacs1999differential}
		R.~Isaacs, {\em Differential games: a mathematical theory with applications to
		  warfare and pursuit, control and optimization}.
		\newblock Courier Corporation, 1999.

		\bibitem{schelling1980strategy}
		T.~C. Schelling, {\em The Strategy of Conflict: with a new Preface by the
		  Author}.
		\newblock Harvard university press, 1980.

		\bibitem{mehta1994focal}
		J.~Mehta, C.~Starmer, and R.~Sugden, ``Focal points in pure coordination games:
		  An experimental investigation,'' {\em Theory and Decision}, vol.~36, no.~2,
		  pp.~163--185, 1994.

		\bibitem{sugden1995theory}
		R.~Sugden, ``A theory of focal points,'' {\em The Economic Journal}, vol.~105,
		  no.~430, pp.~533--550, 1995.

		\bibitem{rijt2019quest}
		J.-W. v.~d. Rijt, ``The quest for a rational explanation: An overview of the
		  development of focal point theory,'' {\em Focal Points in Negotiation},
		  pp.~15--44, 2019.

		\bibitem{czyzowicz2019symmetry}
		J.~Czyzowicz, L.~Gasieniec, R.~Killick, and E.~Kranakis, ``Symmetry breaking in
		  the plane: Rendezvous by robots with unknown attributes,'' in {\em
		  Proceedings of the 2019 ACM Symposium on Principles of Distributed
		  Computing}, pp.~4--13, 2019.

		\bibitem{alpern1995rendezvousAndGal}
		S.~Alpern and S.~Gal, ``Rendezvous search on the line with distinguishable
		  players,'' {\em SIAM Journal on Control and Optimization}, vol.~33, no.~4,
		  pp.~1270--1276, 1995.

		\bibitem{de2006asynchronous}
		G.~De~Marco, L.~Gargano, E.~Kranakis, D.~Krizanc, A.~Pelc, and U.~Vaccaro,
		  ``Asynchronous deterministic rendezvous in graphs,'' {\em Theoretical
		  Computer Science}, vol.~355, no.~3, pp.~315--326, 2006.

		\bibitem{anderson1990rendezvous}
		E.~J. Anderson and R.~Weber, ``The rendezvous problem on discrete locations,''
		  {\em Journal of Applied Probability}, vol.~27, no.~4, pp.~839--851, 1990.

		\bibitem{kranakis2006mobile}
		E.~Kranakis, D.~Krizanc, and S.~Rajsbaum, ``Mobile agent rendezvous: A
		  survey,'' in {\em International Colloquium on Structural Information and
		  Communication Complexity}, pp.~1--9, Springer, 2006.

		\bibitem{pelc2012deterministic}
		A.~Pelc, ``Deterministic rendezvous in networks: A comprehensive survey,'' {\em
		  Networks}, vol.~59, no.~3, pp.~331--347, 2012.

		\bibitem{howard1999rendezvous}
		J.~Howard, ``Rendezvous search on the interval and the circle,'' {\em
		  Operations Research}, vol.~47, no.~4, pp.~550--558, 1999.

		\bibitem{collins2010tell}
		A.~Collins, J.~Czyzowicz, L.~G{\k{a}}sieniec, and A.~Labourel, ``Tell me where
		  i am so i can meet you sooner,'' in {\em International Colloquium on
		  Automata, Languages, and Programming}, pp.~502--514, Springer, 2010.

		\bibitem{collins2011synchronous}
		A.~Collins, J.~Czyzowicz, L.~G{\k{a}}sieniec, A.~Kosowski, and R.~Martin,
		  ``Synchronous rendezvous for location-aware agents,'' in {\em International
		  Symposium on Distributed Computing}, pp.~447--459, Springer, 2011.

		\bibitem{banerjee2020study}
		S.~Banerjee and S.~G. Chaudhuri, ``A study of gathering of location-aware
		  mobile robots,'' in {\em Proceedings of the Global AI Congress 2019},
		  pp.~579--588, Springer, 2020.

		\bibitem{czyzowicz2020gathering}
		J.~Czyzowicz, R.~Killick, E.~Kranakis, D.~Krizanc, and O.~Morales-Ponce,
		  ``Gathering in the plane of location-aware robots in the presence of spies,''
		  {\em Theoretical Computer Science}, vol.~836, pp.~94--109, 2020.

		\bibitem{yu1996agent}
		X.~Yu and M.~Yung, ``Agent rendezvous: A dynamic symmetry-breaking problem,''
		  in {\em International Colloquium on Automata, Languages, and Programming},
		  pp.~610--621, Springer, 1996.

		\bibitem{fraigniaud2008deterministic}
		P.~Fraigniaud and A.~Pelc, ``Deterministic rendezvous in trees with little
		  memory,'' in {\em International Symposium on Distributed Computing},
		  pp.~242--256, Springer, 2008.

		\bibitem{pelc2019using}
		A.~Pelc and R.~N. Yadav, ``Using time to break symmetry: Universal
		  deterministic anonymous rendezvous,'' in {\em The 31st ACM Symposium on
		  Parallelism in Algorithms and Architectures}, pp.~85--92, 2019.

		\bibitem{lim1997rendezvous}
		W.~S. Lim, S.~Alpern, and A.~Beck, ``Rendezvous search on the line with more
		  than two players,'' {\em Operations Research}, vol.~45, no.~3, pp.~357--364,
		  1997.

		\bibitem{lim1996minimax}
		W.~S. Lim and S.~Alpern, ``Minimax rendezvous on the line,'' {\em SIAM Journal
		  on Control and Optimization}, vol.~34, no.~5, pp.~1650--1665, 1996.

		\bibitem{alpern1997rendezvous}
		S.~Alpern and A.~Beck, ``Rendezvous search on the line with bounded resources:
		  expected time minimization,'' {\em European journal of operational research},
		  vol.~101, no.~3, pp.~588--597, 1997.

		\bibitem{alpern1999rendezvous}
		S.~Alpern and A.~Beck, ``Rendezvous search on the line with limited resources:
		  Maximizing the probability of meeting,'' {\em Operations Research}, vol.~47,
		  no.~6, pp.~849--861, 1999.

		\bibitem{alpern2006rendezvous}
		S.~Alpern and V.~Baston, ``Rendezvous in higher dimensions,'' {\em SIAM journal
		  on control and optimization}, vol.~44, no.~6, pp.~2233--2252, 2006.

		\bibitem{dani2016codes}
		V.~Dani, T.~P. Hayes, C.~Moore, and A.~Russell, ``Codes, lower bounds, and
		  phase transitions in the symmetric rendezvous problem,'' {\em Random
		  Structures \& Algorithms}, vol.~49, no.~4, pp.~742--765, 2016.

		\bibitem{georgiou2019symmetric}
		K.~Georgiou, J.~Griffiths, and Y.~Yakubov, ``Symmetric rendezvous with advice:
		  How to rendezvous in a disk,'' {\em Journal of Parallel and Distributed
		  Computing}, vol.~134, pp.~13--24, 2019.

		\bibitem{czyzowicz2012meet}
		J.~Czyzowicz, A.~Kosowski, and A.~Pelc, ``How to meet when you forget:
		  log-space rendezvous in arbitrary graphs,'' {\em Distributed Computing},
		  vol.~25, no.~2, pp.~165--178, 2012.

		\bibitem{fomin2021can}
		F.~V. Fomin, P.~A. Golovach, and D.~M. Thilikos, ``Can romeo and juliet meet?
		  or rendezvous games with adversaries on graphs,'' in {\em International
		  Workshop on Graph-Theoretic Concepts in Computer Science}, pp.~308--320,
		  Springer, 2021.

		\bibitem{lin2003multi}
		J.~Lin, A.~S. Morse, and B.~D. Anderson, ``The multi-agent rendezvous
		  problem,'' in {\em 42nd ieee international conference on decision and control
		  (ieee cat. no. 03ch37475)}, vol.~2, pp.~1508--1513, IEEE, 2003.

		\bibitem{lin2007multi}
		J.~Lin, A.~S. Morse, and B.~D. Anderson, ``The multi-agent rendezvous problem.
		  part 2: The asynchronous case,'' {\em SIAM Journal on Control and
		  Optimization}, vol.~46, no.~6, pp.~2120--2147, 2007.

		\bibitem{cieliebak2003solving}
		M.~Cieliebak, P.~Flocchini, G.~Prencipe, and N.~Santoro, ``Solving the robots
		  gathering problem,'' in {\em International Colloquium on Automata, Languages,
		  and Programming}, pp.~1181--1196, Springer, 2003.

		\bibitem{kranakis2003mobile}
		E.~Kranakis, N.~Santoro, C.~Sawchuk, and D.~Krizanc, ``Mobile agent rendezvous
		  in a ring,'' in {\em 23rd International Conference on Distributed Computing
		  Systems, 2003. Proceedings.}, pp.~592--599, IEEE, 2003.

		\bibitem{baston2001rendezvous}
		V.~Baston and S.~Gal, ``Rendezvous search when marks are left at the starting
		  points,'' {\em Naval Research Logistics (NRL)}, vol.~48, no.~8, pp.~722--731,
		  2001.

		\bibitem{Dirac1947}
		P.~A.~M. Dirac, {\em The Principles of Quantum Mechanics}.
		\newblock Oxford University Press, 1947.

		\bibitem{popescu1994quantum}
		S.~Popescu and D.~Rohrlich, ``Quantum nonlocality as an axiom,'' {\em
		  Foundations of Physics}, vol.~24, no.~3, pp.~379--385, 1994.

		\bibitem{uola2020quantum}
		R.~Uola, A.~C. Costa, H.~C. Nguyen, and O.~G{\"u}hne, ``Quantum steering,''
		  {\em Reviews of Modern Physics}, vol.~92, no.~1, p.~015001, 2020.

		\bibitem{schrodinger1935discussion}
		E.~Schr{\"o}dinger, ``Discussion of probability relations between separated
		  systems,'' in {\em Mathematical Proceedings of the Cambridge Philosophical
		  Society}, vol.~31, pp.~555--563, Cambridge University Press, 1935.

		\bibitem{wiseman2007steering}
		H.~M. Wiseman, S.~J. Jones, and A.~C. Doherty, ``Steering, entanglement,
		  nonlocality, and the einstein-podolsky-rosen paradox,'' {\em Physical review
		  letters}, vol.~98, no.~14, p.~140402, 2007.

		\bibitem{oppenheim2010uncertainty}
		J.~Oppenheim and S.~Wehner, ``The uncertainty principle determines the
		  nonlocality of quantum mechanics,'' {\em Science}, vol.~330, no.~6007,
		  pp.~1072--1074, 2010.

		\bibitem{ramanathan2018steering}
		R.~Ramanathan, D.~Goyeneche, S.~Muhammad, P.~Mironowicz, M.~Gr{\"u}nfeld,
		  M.~Bourennane, and P.~Horodecki, ``Steering is an essential feature of
		  non-locality in quantum theory,'' {\em Nature communications}, vol.~9, no.~1,
		  pp.~1--6, 2018.

		\bibitem{clauser1969proposed}
		J.~F. Clauser, M.~A. Horne, A.~Shimony, and R.~A. Holt, ``Proposed experiment
		  to test local hidden-variable theories,'' {\em Physical review letters},
		  vol.~23, no.~15, p.~880, 1969.

		\bibitem{cirel1980quantum}
		B.~S. Cirel'son, ``Quantum generalizations of bell's inequality,'' {\em Letters
		  in Mathematical Physics}, vol.~4, no.~2, pp.~93--100, 1980.

		\bibitem{gill2003accardi}
		R.~D. Gill, ``Accardi contra bell (cum mundi): The impossible coupling,'' {\em
		  Lecture Notes-Monograph Series}, pp.~133--154, 2003.

		\bibitem{navascues2007bounding}
		M.~Navascu{\'e}s, S.~Pironio, and A.~Ac{\'\i}n, ``Bounding the set of quantum
		  correlations,'' {\em Physical Review Letters}, vol.~98, no.~1, p.~010401,
		  2007.

		\bibitem{navascues2008convergent}
		M.~Navascu{\'e}s, S.~Pironio, and A.~Ac{\'\i}n, ``A convergent hierarchy of
		  semidefinite programs characterizing the set of quantum correlations,'' {\em
		  New Journal of Physics}, vol.~10, no.~7, p.~073013, 2008.

		\bibitem{navascues2010glance}
		M.~Navascu{\'e}s and H.~Wunderlich, ``A glance beyond the quantum model,'' {\em
		  Proceedings of the Royal Society A: Mathematical, Physical and Engineering
		  Sciences}, vol.~466, no.~2115, pp.~881--890, 2010.

		\bibitem{vandenberghe1996semidefinite}
		L.~Vandenberghe and S.~Boyd, ``Semidefinite programming,'' {\em SIAM review},
		  vol.~38, no.~1, pp.~49--95, 1996.

		\bibitem{brunner2014bell}
		N.~Brunner, D.~Cavalcanti, S.~Pironio, V.~Scarani, and S.~Wehner, ``Bell
		  nonlocality,'' {\em Reviews of Modern Physics}, vol.~86, no.~2, p.~419, 2014.

		\bibitem{einstein1935can}
		A.~Einstein, B.~Podolsky, and N.~Rosen, ``Can quantum-mechanical description of
		  physical reality be considered complete?,'' {\em Physical review}, vol.~47,
		  no.~10, p.~777, 1935.

		\bibitem{aspect1982experimental}
		A.~Aspect, J.~Dalibard, and G.~Roger, ``Experimental test of bell's
		  inequalities using time-varying analyzers,'' {\em Physical review letters},
		  vol.~49, no.~25, p.~1804, 1982.

		\bibitem{aspect1982experimental2}
		A.~Aspect, P.~Grangier, and G.~Roger, ``Experimental realization of
		  einstein-podolsky-rosen-bohm gedankenexperiment: a new violation of bell's
		  inequalities,'' {\em Physical review letters}, vol.~49, no.~2, p.~91, 1982.

		\bibitem{hensen2015loophole}
		B.~Hensen, H.~Bernien, A.~E. Dr{\'e}au, A.~Reiserer, N.~Kalb, M.~S. Blok,
		  J.~Ruitenberg, R.~F. Vermeulen, R.~N. Schouten, C.~Abell{\'a}n, {\em et~al.},
		  ``Loophole-free bell inequality violation using electron spins separated by
		  1.3 kilometres,'' {\em Nature}, vol.~526, no.~7575, pp.~682--686, 2015.

		\bibitem{giustina2015significant}
		M.~Giustina, M.~A. Versteegh, S.~Wengerowsky, J.~Handsteiner, A.~Hochrainer,
		  K.~Phelan, F.~Steinlechner, J.~Kofler, J.-{\AA}. Larsson, C.~Abell{\'a}n,
		  {\em et~al.}, ``Significant-loophole-free test of bell’s theorem with
		  entangled photons,'' {\em Physical review letters}, vol.~115, no.~25,
		  p.~250401, 2015.

		\bibitem{yin2017satellite}
		J.~Yin, Y.~Cao, Y.-H. Li, S.-K. Liao, L.~Zhang, J.-G. Ren, W.-Q. Cai, W.-Y.
		  Liu, B.~Li, H.~Dai, {\em et~al.}, ``Satellite-based entanglement distribution
		  over 1200 kilometers,'' {\em Science}, vol.~356, no.~6343, pp.~1140--1144,
		  2017.

		\bibitem{brassard1999cost}
		G.~Brassard, R.~Cleve, and A.~Tapp, ``Cost of exactly simulating quantum
		  entanglement with classical communication,'' {\em Physical Review Letters},
		  vol.~83, no.~9, p.~1874, 1999.

		\bibitem{brassard2005quantum}
		G.~Brassard, A.~Broadbent, and A.~Tapp, ``Quantum pseudo-telepathy,'' {\em
		  Foundations of Physics}, vol.~35, no.~11, pp.~1877--1907, 2005.

		\bibitem{brukner2006entanglement}
		{\v{C}}.~Brukner, N.~Paunkovi{\'c}, T.~Rudolph, and V.~Vedral,
		  ``Entanglement-assisted orientation in space,'' {\em International Journal of
		  Quantum Information}, vol.~4, no.~02, pp.~365--370, 2006.

		\bibitem{summhammer2005quantum}
		J.~Summhammer, ``Quantum cooperation of two insects,'' {\em arXiv preprint
		  quant-ph/0503136}, 2005.

		\bibitem{summhammer2011quantum}
		J.~Summhammer, ``Quantum cooperation,'' {\em Axiomathes}, vol.~21, no.~2,
		  pp.~347--356, 2011.

		\bibitem{nakayama2010experimental}
		S.~Nakayama and I.~Iimura, ``Experimental study on quantum-entangled
		  cooperative behavior of two ants,'' in {\em 2010 Second World Congress on
		  Nature and Biologically Inspired Computing (NaBIC)}, pp.~566--571, IEEE,
		  2010.

		\bibitem{iimura2014experimental}
		I.~Iimura, K.~Takeda, and S.~Nakayama, ``An experimental study on
		  quantum-entangled cooperative behavior in swarm intelligence,'' {\em
		  International Journal of Emerging Technology and Advanced Engineering},
		  vol.~4, no.~8, 2014.

		\bibitem{iimura2015effect}
		I.~Iimura, K.~Takeda, and S.~Nakayama, ``Effect of quantum cooperation in three
		  entangled ants,'' {\em International Journal of Emerging Technology and
		  Advanced Engineering}, vol.~5, no.~8, 2015.

		\bibitem{khoshnoud2017quantum}
		F.~Khoshnoud, C.~W. de~Silva, and I.~I. Esat, ``Quantum entanglement of
		  autonomous vehicles for cyber-physical security,'' in {\em 2017 IEEE
		  International Conference on Systems, Man, and Cybernetics (SMC)},
		  pp.~2655--2660, IEEE, 2017.

		\bibitem{khoshnoud2019quantum}
		F.~Khoshnoud, I.~I. Esat, C.~W. de~Silva, and M.~B. Quadrelli, ``Quantum
		  network of cooperative unmanned autonomous systems,'' {\em Unmanned Systems},
		  vol.~7, no.~02, pp.~137--145, 2019.

		\bibitem{bennet1984quantum}
		C.~BENNET, ``Quantum cryptography: Public key distribution and coin tossing,''
		  in {\em Proc. of IEEE Int. Conf. on Comp. Sys. and Signal Proc., Dec. 1984},
		  1984.

		\bibitem{khoshnoud2020quantum}
		F.~Khoshnoud, M.~B. Quadrelli, I.~I. Esat, and D.~Robinson, ``Quantum
		  cooperative robotics and autonomy,'' {\em arXiv preprint arXiv:2008.12230},
		  2020.

		\bibitem{khoshnoud2020quantum2}
		F.~Khoshnoud, I.~I. Esat, S.~Javaherian, and B.~Bahr, ``Quantum entanglement
		  and cryptography for automation and control of dynamic systems,'' {\em arXiv
		  preprint arXiv:2007.08567}, 2020.

		\bibitem{khoshnoud2020quantum3}
		F.~Khoshnoud, L.~Lamata, C.~W. de~Silva, and M.~B. Quadrelli, ``Quantum
		  teleportation for control of dynamic systems and autonomy,'' {\em arXiv
		  preprint arXiv:2007.15249}, 2020.

		\bibitem{bennett1993teleporting}
		C.~H. Bennett, G.~Brassard, C.~Cr{\'e}peau, R.~Jozsa, A.~Peres, and W.~K.
		  Wootters, ``Teleporting an unknown quantum state via dual classical and
		  einstein-podolsky-rosen channels,'' {\em Physical review letters}, vol.~70,
		  no.~13, p.~1895, 1993.

		\bibitem{kumar2021survey}
		A.~Kumar, S.~Bhatia, K.~Kaushik, S.~M. Gandhi, S.~G. Devi, A.~D.~J. Diego, and
		  A.~Mashat, ``Survey of promising technologies for quantum drones and
		  networks,'' {\em IEEE Access}, vol.~9, pp.~125868--125911, 2021.

		\bibitem{pal2010maximal}
		K.~F. P{\'a}l and T.~V{\'e}rtesi, ``Maximal violation of a bipartite
		  three-setting, two-outcome bell inequality using infinite-dimensional quantum
		  systems,'' {\em Physical Review A}, vol.~82, no.~2, p.~022116, 2010.

		\bibitem{colbeck2009quantum}
		R.~Colbeck, ``Quantum and relativistic protocols for secure multi-party
		  computation,'' {\em arXiv preprint arXiv:0911.3814}, 2009.

		\bibitem{ahlswede1987search}
		R.~Ahlswede and I.~Wegener, {\em Search problems}.
		\newblock John Wiley \& Sons, Inc., 1987.

		\bibitem{baston2019search}
		V.~Baston and K.~Kikuta, ``A search problem on a bipartite network,'' {\em
		  European Journal of Operational Research}, vol.~277, no.~1, pp.~227--237,
		  2019.

		\bibitem{elitzur1993quantum}
		A.~C. Elitzur and L.~Vaidman, ``Quantum mechanical interaction-free
		  measurements,'' {\em Foundations of Physics}, vol.~23, no.~7, pp.~987--997,
		  1993.

		\bibitem{kwiat1995interaction}
		P.~Kwiat, H.~Weinfurter, T.~Herzog, A.~Zeilinger, and M.~A. Kasevich,
		  ``Interaction-free measurement,'' {\em Physical Review Letters}, vol.~74,
		  no.~24, p.~4763, 1995.

		\bibitem{OCTAVE}
		J.~W. Eaton, D.~Bateman, S.~Hauberg, and R.~Wehbring, ``Gnu octave manual: A
		  high-level interactive language for numerical computations,'' {\em Network
		  Theory Ltd}, 2007.

		\bibitem{sturm1999using}
		J.~F. Sturm, ``Using sedumi 1.02, a matlab toolbox for optimization over
		  symmetric cones,'' {\em Optimization methods and software}, vol.~11, no.~1-4,
		  pp.~625--653, 1999.

		\bibitem{toh1999sdpt3}
		K.-C. Toh, M.~J. Todd, and R.~H. T{\"u}t{\"u}nc{\"u}, ``Sdpt3—a matlab
		  software package for semidefinite programming, version 1.3,'' {\em
		  Optimization methods and software}, vol.~11, no.~1-4, pp.~545--581, 1999.

		\bibitem{lofberg2004yalmip}
		J.~Lofberg, ``Yalmip: A toolbox for modeling and optimization in matlab,'' in
		  {\em 2004 IEEE international conference on robotics and automation (IEEE Cat.
		  No. 04CH37508)}, pp.~284--289, IEEE, 2004.

		\end{thebibliography}
\end{document}